\RequirePackage{fix-cm}
\documentclass[smallextended]{svjour3}
\smartqed  
\usepackage{graphicx}
\usepackage{caption}
\usepackage{nicefrac}
\usepackage{amsfonts}
\usepackage[T1]{fontenc}
\usepackage{xspace}
\usepackage{flushend}
\usepackage{ifthen}
\usepackage{color}
\usepackage{url}
\usepackage{multirow}
\usepackage{listliketab}
\usepackage{amssymb}
\usepackage[linesnumbered,lined,boxed,commentsnumbered]{algorithm2e}
\usepackage{algorithmic}
\usepackage{booktabs}
\usepackage{comment}
\usepackage{amsmath}
\usepackage{bbding}
\usepackage{listings}
\usepackage{float}
\usepackage[misc]{ifsym}
\usepackage{tabularx}
\usepackage{rotating}
\usepackage{xcolor}
\usepackage{natbib}
\usepackage{tcolorbox}
\usepackage{ulem}
\usepackage{censor}
\usepackage{subcaption}
\usepackage{pgfplots}
\pgfplotsset{compat=1.14}
\usepackage{pgfplotstable}

\PassOptionsToPackage{hyphens}{url}
\usepackage[hidelinks]{hyperref}
\usepackage{tablefootnote}

\definecolor{Fork}{HTML}{99CCFF}
\definecolor{Non-Fork}{HTML}{FFCCCB}

\newcommand\RqThree{RQ1}
\newcommand\RqFour{RQ2}

\newcommand\RqThreeText{(\RqThree) \textit{To what extent do \ExPR s ease maintainer workload by fixing existing issues?}}
\newcommand\RqFourText{(\RqFour) \textit{How different are the code changes when comparing \ExPR s to \InPR s?}}

\newcommand\RqOneResult{Findings indicate that \ExPR s are indeed prevalent with a large number being received by NPM packages (75.02\% on average).
Furthermore, NPM contributors have a high (88.87\% on average) rate of submitting \ExPR s.
We find that on average, \ExPR s' (55.65\% per package)~acceptance is higher than \InPR s' (51.04\% per package).
}

\newcommand\RqThreeResult{
We find that 26.75\% of \ExPR s submitted to a package are linked to an already existing issue.
Furthermore, we find evidence that there are \ExPR s that require the attention of the core team (e.g., referencing breaking changes, urgent, and on-hold labels).}
\newcommand\RqFourResult{Similar to other PR types (internal and bots), \ExPR s are most likely to contain changes to add new features (170 out of 384 PRs) to a package.
However, unlike Internal and Bot PRs, \ExPR s tend to focus on documentation (44 out of 384) as opposed to features (380 out of 384) for bots, and refactoring (34 out of 384) for \InPR s.}

\newcommand\ExPR{External PR}
\newcommand\InPR{Internal PR}
\newcommand\BotPR{Bot PR}

\newcommand\ExRatePck{Ext-PRrate (pkg)}
\newcommand\ExRateCont{Ext-PRrate (contrib.)}

\newcommand\AcPR{Accepted PR}
\newcommand\AbPR{Abandoned PR}

\newcommand\HNullTwoOne{$H2.1_{null}$ \textit{There is no difference between the acceptance and abandonment of \ExPR s}}
\newcommand\HNullTwoTwo{$H2.2_{null}$ \textit{There is no difference in PR acceptance between external, internal or bot PRs}}

\usepackage{tikz}

\colorlet{punct}{red!60!black}
\definecolor{background}{HTML}{EEEEEE}
\definecolor{delim}{RGB}{20,105,176}
\colorlet{numb}{magenta!60!black}

\lstdefinelanguage{json}{
    basicstyle=\small,
    numbers=left,
    numberstyle=\scriptsize,
    stepnumber=1,
    numbersep=8pt,
    showstringspaces=false,
    breaklines=true,
    frame=lines,
    backgroundcolor=\color{background},
    literate=
     *{0}{{{\color{numb}0}}}{1}
      {1}{{{\color{numb}1}}}{1}
      {2}{{{\color{numb}2}}}{1}
      {3}{{{\color{numb}3}}}{1}
      {4}{{{\color{numb}4}}}{1}
      {5}{{{\color{numb}5}}}{1}
      {6}{{{\color{numb}6}}}{1}
      {7}{{{\color{numb}7}}}{1}
      {8}{{{\color{numb}8}}}{1}
      {9}{{{\color{numb}9}}}{1}
      {:}{{{\color{punct}{:}}}}{1}
      {,}{{{\color{punct}{,}}}}{1}
      {\{}{{{\color{delim}{\{}}}}{1}
      {\}}{{{\color{delim}{\}}}}}{1}
      {[}{{{\color{delim}{[}}}}{1}
      {]}{{{\color{delim}{]}}}}{1},
}

\begin{document}

\title{Understanding the Role of External Pull Requests in the NPM Ecosystem}


\author{Vittunyuta Maeprasart \Letter\and
        Supatsara Wattanakriengkrai\and
        Raula Gaikovina Kula\and
        Christoph Treude\and
        Kenichi Matsumoto
}

\institute{
    \Letter~Corresponding author - Vittunyuta Maeprasart \and Supatsara Wattanakriengkrai \and Raula Gaikovina Kula \and Kenichi Matsumoto
    \at Nara Institute of Science and Technology, Japan \\
    \email{\{maeprasart.vittunyuta.mn2, wattanakri.supatsara.ws3, raula-k, matumoto\}@is.naist.jp} \\
    Christoph Treude \at University of Melbourne, Australia \\ \email{ctreude@gmail.com} \and
}

\date{Received: date / Accepted: date}

\maketitle

\begin{abstract}
The risk to using third-party libraries in a software application is that much needed maintenance is solely carried out by library maintainers. 
These libraries may rely on a core team of maintainers (who might be a single maintainer that is unpaid and overworked) to serve a massive client user-base.
On the other hand, being open source has the benefit of receiving contributions (in the form of \ExPR s) to help fix bugs and add new features. 
In this paper, we investigate the role by which \ExPR s~(contributions from outside the core team of maintainers) contribute to a library.
Through a preliminary analysis, we find that \ExPR s~are prevalent, and just as likely to be accepted as maintainer PRs.
We find that 26.75\% of \ExPR s submitted fix existing issues.
Moreover, fixes also belong to labels such as \texttt{breaking changes}, \texttt{urgent}, and \texttt{on-hold}.
Differently from \InPR s, \ExPR s cover documentation changes (44 out of 384 PRs), while not having as much refactoring (34 out of 384 PRs). On the other hand, \ExPR s~also cover new features (380 out of 384 PRs) and bugs (120 out of 384).
Our results lay the groundwork for understanding how maintainers decide which external contributions they select to evolve their libraries and what role they play in reducing the workload.
\end{abstract}

\keywords{ Third-party libraries, Pull Requests, OSS sustainability, Software Ecosystems}

\section{Introduction and Motivation}
\label{sec:introduction}
Third-party libraries provide a means by which software teams can quickly build an application, avoiding the need to start from scratch.
Popular usage has led to the explosion of third-party library inter-connecting networks of dependencies that form large ecosystems \citep{IslamICSME2021,decan2018impact,HeHao2021}.
For example, the NPM ecosystem of libraries (aka packages) is a critical part of the JavaScript world, with open source contributions from hundreds of thousands of open source developers and maintainers.
Evidence of its impact was its purchase by Microsoft's GitHub in early 2020 \citep{npmisjoi86_online}. 
Despite their popularity, libraries are prone to maintenance and evolution issues. 
In particular, threats such as vulnerabilities \citep{chinthanet2021lags, Durumeric2014Heartbleed} and transitive dependency changes \citep{dey2019_promise_mockus} may cause a library to risk becoming obsolete \citep{kula2018developers}, or worse, pose an important security risk. 

Recent events such as the Log4Shell vulnerability \citep{Log4Shell_online, Log4jvul_online} and the faker sabotage \citep{npmLibra57_online} 
illustrate concerns about how maintainers become overwhelmed with their workload.
This is especially the case when there is a small core maintainer team (sometimes a single maintainer) that is depended upon by a massive user base.
For instance, it was reported that \textit{`a developer of two popular libraries, decided to inject malicious code into them, citing that the reason behind this mischief on the developer's part appears to be retaliation—against mega-corporations and commercial consumers of open-source projects who extensively rely on cost-free and community-powered software but do not, according to the developer, give back to the community'} \citep{OSScorrupts_online}.
Another example is when the maintainers of a vulnerable library expressed their frustration when maintaining a library in a tweet, \textit{`... maintainers have been working sleeplessly on mitigation measures; fixes, docs, CVE, replies to inquiries, etc. Yet nothing is stopping people to bash us, for work we aren't paid for, for a feature we all dislike yet needed to keep due to backward compatibility concerns...'} \citep{twitter_online}.

Recent initiatives such as the Alpha-Omega project allow industry (Google and Microsoft) to partner with maintainers to systematically find and fix vulnerabilities that have not yet been discovered in open source code \citep{AlphaOmega_article_online}.
Although these critical contributions are indeed significant, little is known about the other external contributions that these libraries constantly receive from the ecosystem. 
Although prior work shows that external contributions are needed, analysis is at a higher level of granularity. 
For instance, \citet{Samoladas2010survival} show that OSS projects require a constant stream of contributions, while \citet{Nakakoji2003} and \citet{raymond1999cathedral} depict external contributions as more casual contributions.
In this light, it is unknown how external contributions ease maintainer workload, and how they differ from maintainer contributions (i.e., fixing bugs, new features, documentation, etc.).

\begin{figure*}[!]
    \centering
    \includegraphics[width=1\linewidth]{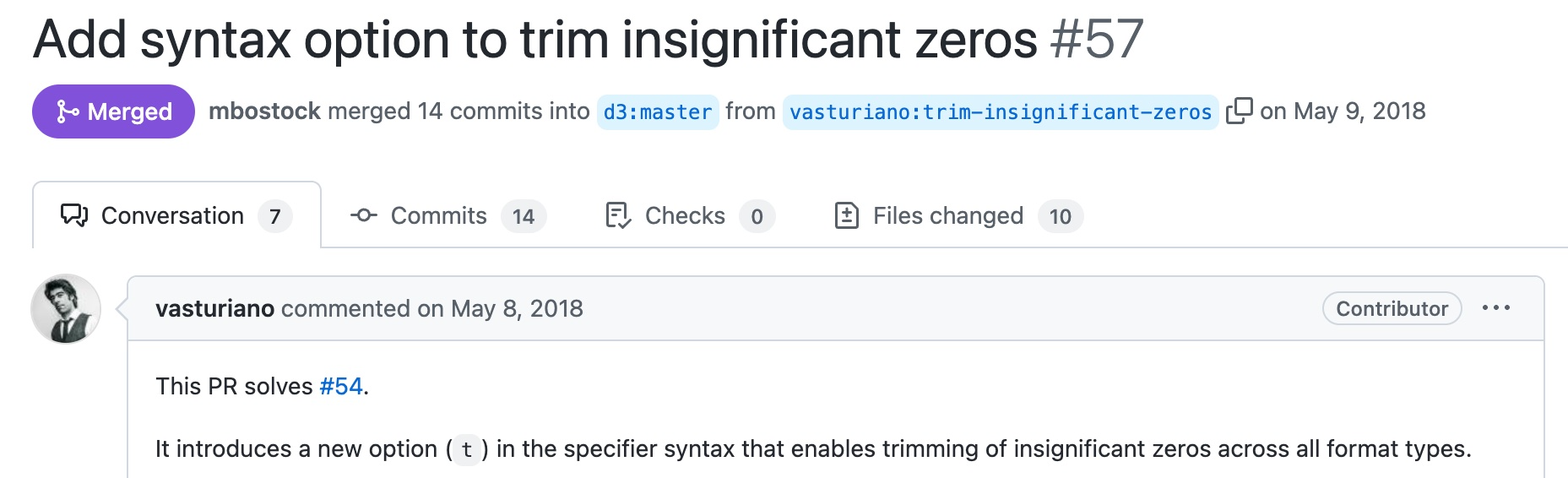}
    \caption{An \ExPR~which is linked to an issue using the keyword \textit{solves}.}
    \label{fig:external_pr}
\end{figure*}

In this study, we analyze contributions submitted as Pull Requests (PRs) in the NPM ecosystem.
Based on prior work \citep{Valiev2018sustained}, we classify a PR contribution based on the access level (i.e., can merge a PR into the code base) of the submitter.
In other words, we define an {\ExPR} as a PR submitted by a contributor who is not a maintainer, such as the example shown in Figure \ref{fig:external_pr} where a contributor submits a PR that solves an already existing issue.
In contrast, an {\InPR} is a PR submitted by a maintainer. Finally, a {\BotPR} is a PR submitted by an automated bot such as dependabot.\footnote{\url{https://github.com/dependabot}}
We collect 1,076,123 PRs from 47,959 NPM packages to first conduct a preliminary study.
The results indicated that \ExPR s are indeed prevalent with a large number being received by a NPM package (75.02\% on average).
Furthermore, NPM contributors have a high (88.87\% on average) rate of submitting \ExPR s.
We find that on average, \ExPR s' (55.65\% per package)~acceptance is higher than \InPR s' (51.04\% per package).
Encouraged by these results, and to further explore the role of \ExPR s, we carried out an empirical study to answer these two research questions:
\begin{itemize}
\item \textbf{\RqThreeText} 
Our motivation for this research question is to confirm the extent to which \ExPR s fix an issue that is already part of the maintainers' workload.

\item \textbf{\RqFourText}
For a qualitative understanding of \ExPR s, for this RQ, we extract and compare \ExPR s with \InPR s. 
\end{itemize}

The results of RQ1 show that 26.75\% of \ExPR s submitted to a package are linked to an already existing issue.
Furthermore, we find evidence that there are \ExPR s that require the attention of the core team (e.g., referencing breaking changes, urgent, and on-hold labels).
For RQ2, similar to other PR types (internal and bots), we find that \ExPR s are most likely to contain changes to add new features (170 out of 384 PRs) to a package.
However, different to Internal and Bot PRs, \ExPR s tend to focus on documentation (44 out of 384) as opposed to features (380 out of 384) for bots, and refactoring (34 out of 384) for \InPR s.

The rest of the paper is organized as follows. 
Section \ref{sec:data_preparation_approach} provides the data preparation processes for our datasets and the preliminary study. Section \ref{sec:empirical_study} provides approach and findings for each RQ. Section \ref{sec:recommendations} discusses lesson learnt from our work. Section \ref{sec:threats_to_validity} exposes potential threats to validity. Section \ref{sec:related_work} reviews previous studies related to our study. Section \ref{sec:conclusion} concludes the paper.
We provide a replication dataset that includes all our scripts, tools and datasets for researchers to use in the future at: \url{https://github.com/NAIST-SE/External-PullRequest}.

\section{Data Preparation and Preliminary Study}
\label{sec:data_preparation_approach}

 In this section, we present our data preparation and preliminary study. 
 
\subsection{Data Collection}
\begin{figure*}[!]
    \centering
    \includegraphics[width=\linewidth]{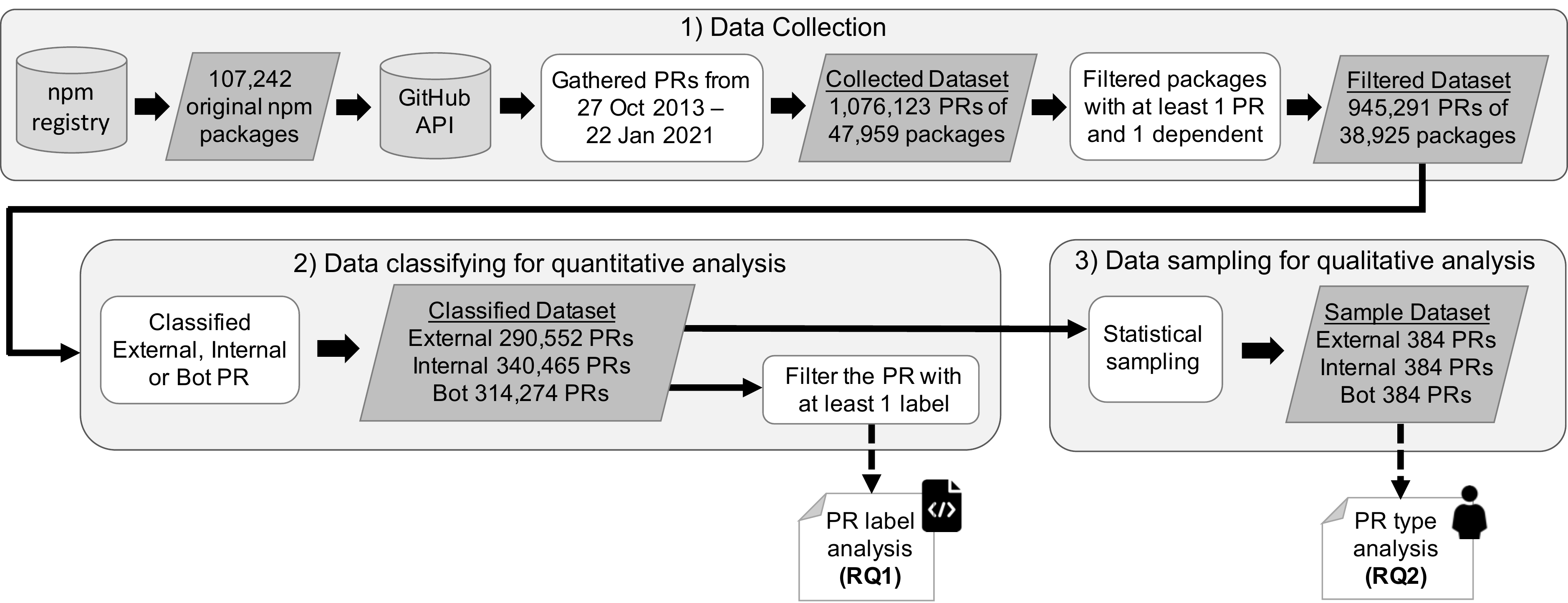}
    \caption{Overview of the data preparation. It consists of three stages: (i) Data collection, (ii) Data classifying for quantitative analysis for \RqThree, and (iii) Data sampling and qualitative analysis for \RqFour}
    \label{fig:data_flow}
\end{figure*}

As shown in Figure \ref{fig:data_flow}, we separate our data collection process into three stages.
\paragraph{\textit{Stage One - Data Collection.}}
For this study, we have two data sources to construct our dataset.
The first data source is the NPM registry which contains a listing of the available NPM packages.
We select the NPM packages due to their popularity (cf. Section \ref{sec:introduction}) and having been the subject of many studies \citep{chinthanet2021lags, Cogo2019, decan2018impact, abdalkareem2017developers, 2018_technique_lag, dey2019_promise_mockus, Dey:ESEM2020}.
Following related work, we obtained this listing following the method of \citet{chinthanet2021lags}, ending up with 107,242 original NPM packages.
The second data source is for obtaining the PR information.
For this, we use the GitHub API\footnote{\url{https://docs.github.com/en/rest/reference/pulls\#get-a-pull-request}}.
We obtained 1,076,123 PRs from 47,959 packages.
Note that we filter out packages that do not have any PR (i.e., 8,991 packages with no dependencies and 67 packages had no PRs submitted), to ensure that our dataset contains active packages with PRs.
The collected PRs contain the PR information, e.g., PR title, description, status, label, and code patches.

To ensure a quality dataset, we further filter out libraries that have no other library depending on them. 
This filter ensures that each package meets the minimum requirements for our analysis.
In the end, we obtained 945,921 PRs from 38,925 packages.
Data collection was performed in January 2021.
The summary statistics of the collection process are shown in Table \ref{tab:dataset_statistic}.

\paragraph{\textit{Stage Two - Data classifying for quantitative analysis (Classified Dataset).}} 
In this stage, we classify whether a PR is an \ExPR, an \InPR, or a \BotPR.
As shown in Table \ref{tab:dataset_statistic}, by identifying the 116,265 contributors to the 945,291 PRs, we find that {290,552 (i.e., 30.74\%)} of the filtered PRs are \ExPR s, while {340,465 or 36.02\%} of the filtered PRs are \InPR s.
To identify \BotPR s, in addition to parameters from the GitHub REST API, we follow prior work proposed by \citet{dey2020detecting} and \citet{golzadeh2020bot}, to collect a list of bots provided by GitHub. 
After performing this classification as shown in Table \ref{tab:dataset_statistic}, we identified 314,274 \BotPR s from the filtered dataset. 
A summary of the detected bots is shown in Table \ref{tab:top_bot}.

\begin{table}[]
\centering
\caption{Statistics of the collected dataset and filtered dataset (after stage 1: Data collection), and classified dataset (after stage 2: Quantitative classifying)}
\label{tab:dataset_statistic}
\scalebox{1}{
\begin{tabular}{@{}lr@{}}
\toprule 
\multicolumn{2}{c}{\textbf{Collected Dataset}} \\ \midrule
Repository snapshot &   27 Oct 2013 --- 22 Jan 2021 \\
\# original NPM packages & 107,242 \\
\# collected NPM Packages & 47,959 \\
\# NPM Pull Requests & 1,076,123 \\ \midrule
\multicolumn{2}{c}{\textbf{After Filtering Process}} \\ \midrule
\# NPM Packages & 38,925 \\
\# NPM Pull Requests & 945,291 \\
\# contributors & 116,265 \\ \midrule
\multicolumn{2}{c}{\textbf{Classified Dataset (\RqThree)}} \\ \midrule
\# \BotPR & 314,274  (33.24\%) \\ 
\# \ExPR & 290,552  (30.74\%) \\
\# \InPR & 340,465  (36.02\%) \\ \midrule
\multicolumn{2}{c}{\textbf{Sample Dataset (\RqFour)}} \\ \midrule
\# Pull Requests & 1,152\\
\bottomrule
\end{tabular}
}
\end{table}

\begin{table}[]
\centering
\caption{Summary of Bots detected}
\label{tab:top_bot}
\scalebox{1}{
\begin{tabular}{l|r}
\toprule
\multicolumn{1}{c|}{\textbf{bot name}} & \multicolumn{1}{c}{\textbf{\# PR}} 
\\ \hline
greenkeeperio-bot & 114,945 (36.57\%) 
\\
dependabot-preview{[}bot{]} & 67,933 (21.62\%)  
\\
greenkeeper{[}bot{]} & 62,837 (19.99\%) 
\\
renovate{[}bot{]} & 36,796 (11.71\%) 
\\
dependabot{[}bot{]} & 25,339 (8.06\%) 
\\
others & 6,424 (2.05\%) 
\\ \hline
\multicolumn{1}{r|}{Total} & 314,274 
\\ 
\bottomrule
\end{tabular}}
\end{table}

\paragraph{\textit{Stage 3. Data sampling for qualitative analysis (Sample Dataset).}} 
From the classified dataset obtained in Stage 2, we draw a statistically representative random sample of 384 \ExPR s, 384 \InPR s, and 384 \BotPR s.
These sample sizes allow us to generalize the conclusions about the ratio of PRs with a specific characteristic to all studied PRs with a confidence level of 95\% and a confidence interval of 5\%, as suggested by a prior work \cite{Hata:icse2019} and using the calculator.\footnote{\url{https://www.surveysystem.com/sscalc.htm}}

\subsection{Preliminary study}
To understand the impact of \ExPR, we first want to explore the extent to which NPM packages receive PRs from external contributors, and how many \ExPR s are submitted by contributors.
We then study the extent to which library maintainers accept \ExPR s.
Hence, we ask the following three preliminary questions.

\paragraph{\textbf{(A1) From a Package Perspective}}
\textit{Approach:} We first identify the proportion of \ExPR s in relation to all submitted PRs.
We define the proportion of the \ExPR s of a package and submitted by a contributor as follows. 
For an NPM package \texttt{pkg} and a contributor \texttt{contrib.} we calculate the proportion as:
\begin{itemize}
     \item \texttt{\ExRatePck}:~ $\frac{\# \ExPR}{\# all-PR}$ submitted to a \texttt{pkg}. In this case, a higher proportion means that there are more \ExPR s received by that package. 
\end{itemize} 
where all-PR is the sum of PRs (\ExPR s, \InPR s, \BotPR s) per package.
For statistical validation, we apply Spearman's rank correlation coefficient or Spearman's $\rho$, a non-parametric measure of rank correlation, to analyze the correlations between the number of \ExPR s.
Spearman's $\rho$ value ranges from -1 to 1. 
We analyze as follows: (1) |$\rho$| $<$ 0.20 is Negligible or no relationship,
(2) 0.20 $\leq$ |$\rho$| $<$ 0.30 is Weak positive/negative relationship, 
(3) 0.30 $\leq$ |$\rho$| $<$ 0.40 is Moderate positive/negative relationship, 
(4) 0.40 $\leq$ |$\rho$| $<$ 0.70 is Strong positive/negative relationship,
(5) 0.70 $\leq$ |$\rho$| $<$ 1 is Very strong positive/negative relationship, or
(6) |$\rho$| = 1 is Perfect relationship.

\begin{figure}[]
\centering
    \begin{subfigure}[]{0.4\textwidth}
        \centering
        \includegraphics[width=0.8\linewidth]{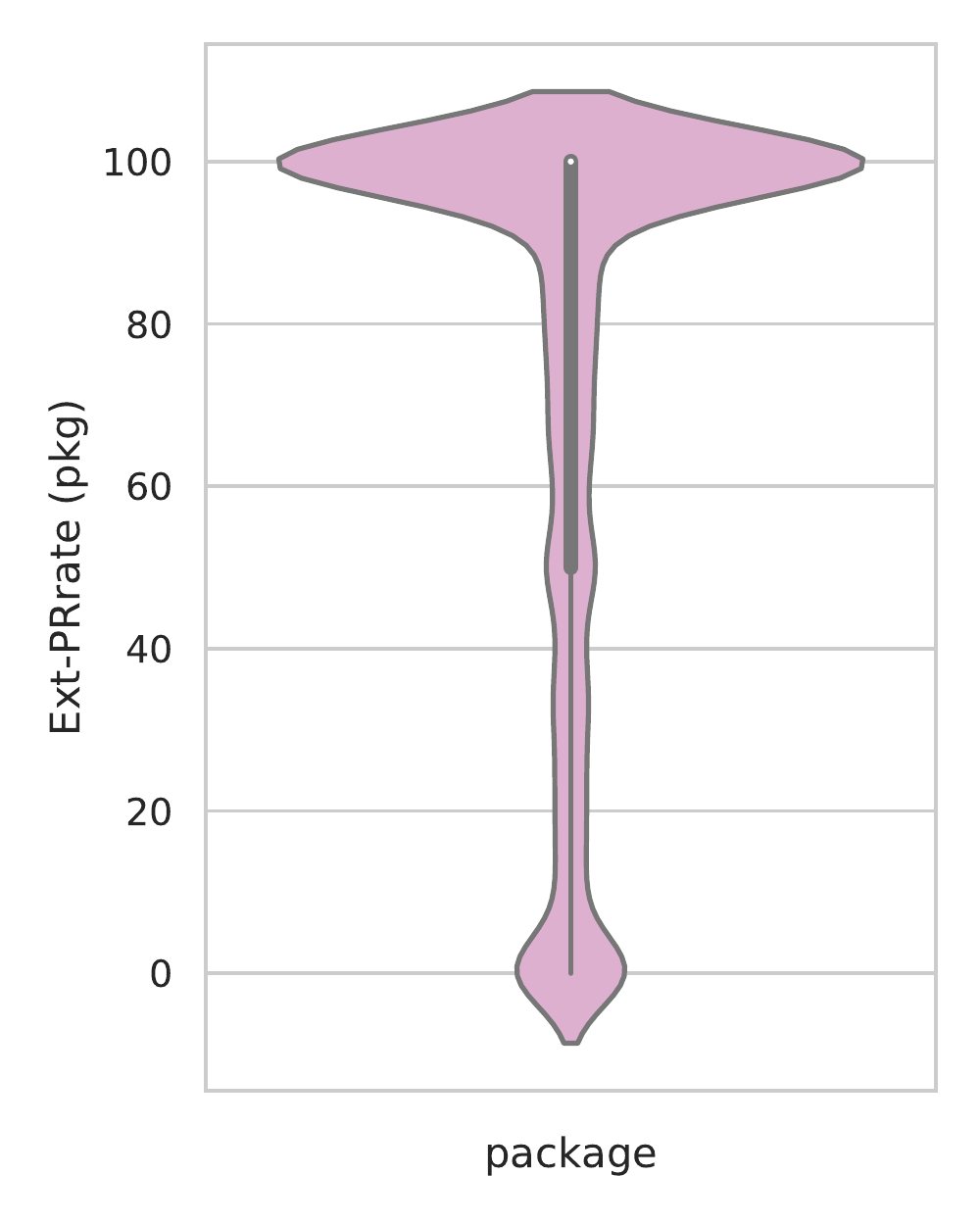}
        \caption{The proportion of \ExPR s a package received (\ExRatePck), i.e., mean of 75.02\%}
        \label{fig:rq1_package_violin}
    \end{subfigure}
    \hfill
    \begin{subfigure}[]{0.4\textwidth}
        \centering
        \includegraphics[width=0.8\linewidth]{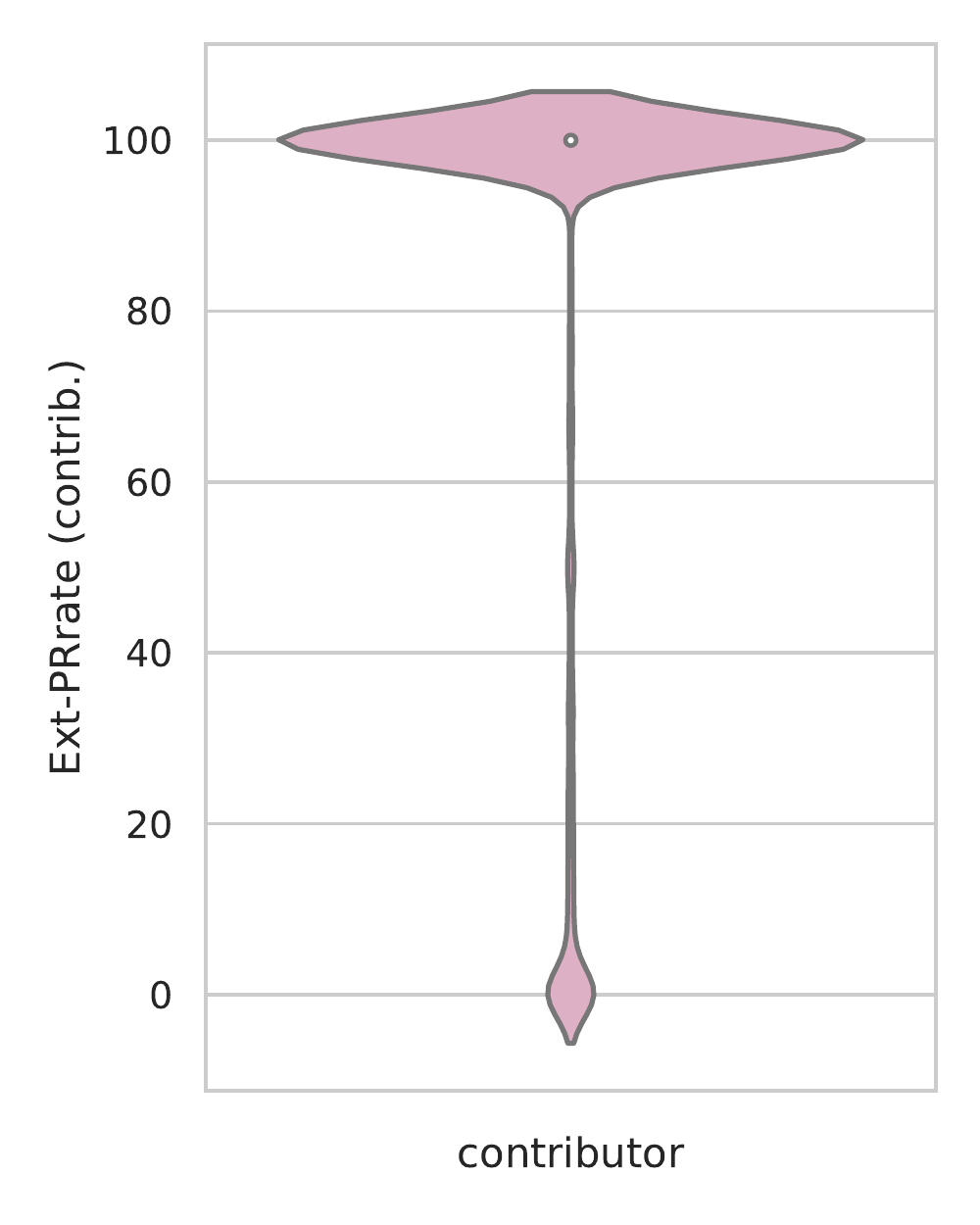}
        \caption{The proportion of \ExPR s a contributor submitted (\ExRateCont), i.e., mean of 88.87\%}
        \label{fig:rq1_contributor_violin}
    \end{subfigure}
    \caption{Prevalent of Packages Receiving an \ExPR~showing in (a) and Contributor submission an \ExPR~showing in (b).}
    \label{fig:rq1_package}
\end{figure}

\textit{Results:}
From the perspective of a package, we discuss two findings.
The first finding is that, from Figure \ref{fig:rq1_package}a, we show visually that \ExPR s are indeed prevalent, as shown by the high \ExRatePck.
The findings show that on average, 75.02\% of PRs received by NPM packages are \ExPR s.

\paragraph{\textbf{(A2) From a Contributor Perspective}}
\textit{Approach:} Similar to A1 we identify the proportion of \ExPR s in relation to all submitted PRs.
However, we now define the proportion of \ExPR s submitted by a contributor as follows. 
For an NPM package \texttt{pkg} and a contributor \texttt{contrib.} we calculate the proportion as:
 \begin{itemize} 
     \item \texttt{\ExRateCont}:~ $\frac{\# \ExPR}{\# all-PR}$ submitted by a \texttt{contrib}. In this case, a higher proportion means that there are more \ExPR s submitted by a contributor. Our statistical validation is the same as P1.
\end{itemize} 

\textit{Results:}
Similar to the package perspective, we discuss two findings for the analysis of contributor submissions of \ExPR s.
The first finding is that, from Figure \ref{fig:rq1_package}b, we confirm visually that a high percentage of contributions submitted by a contributor are \ExPR s. 
This is evident by an average of 88.87\% being \ExPR s per contributor.

\begin{figure}[!]
    \centering
    \includegraphics[width=0.8\linewidth]{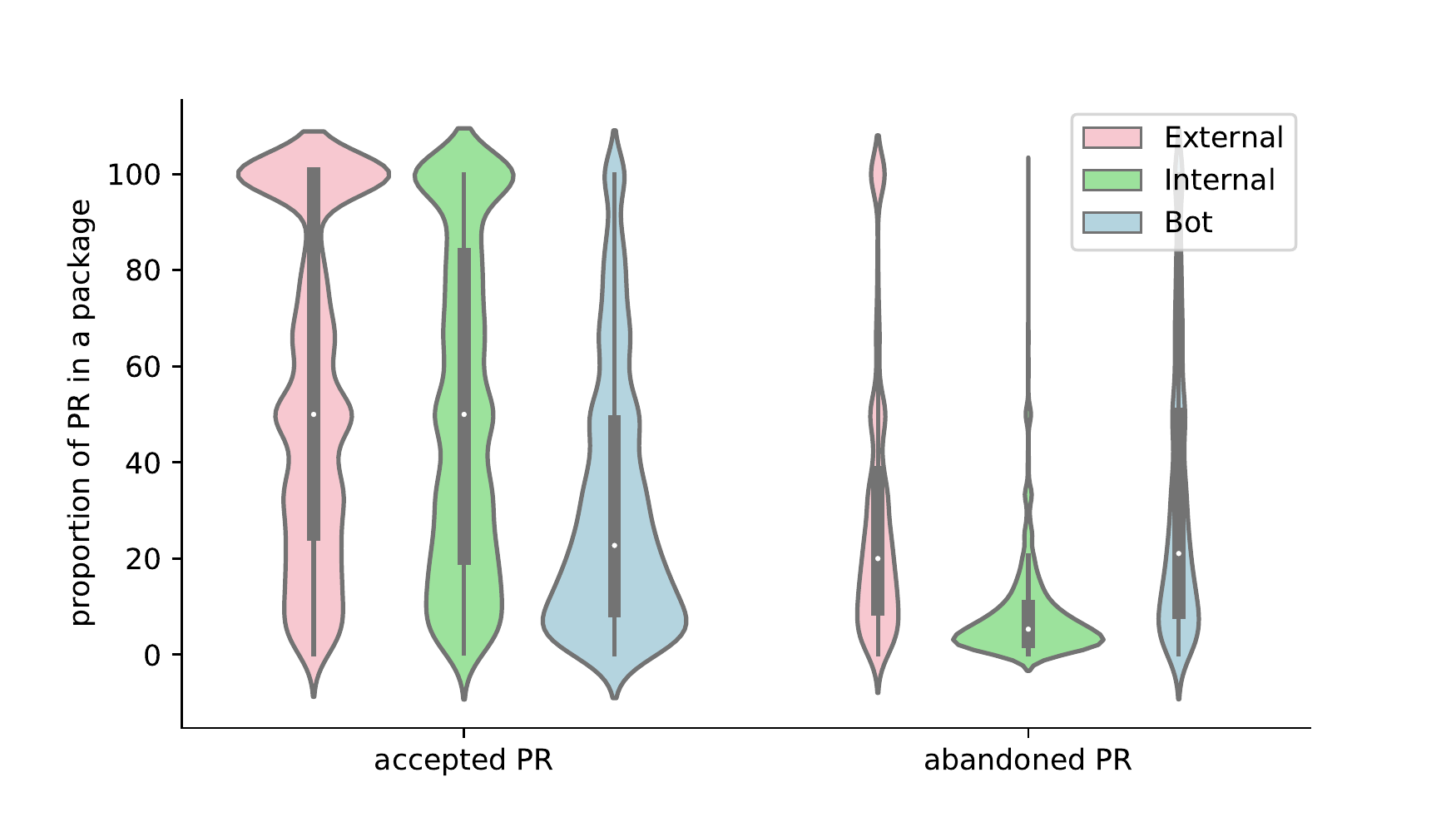}
    \caption{Distributions of PR state by the type of PR. Note that External (Pink), Internal (Green), and Bot (Blue) PRs are shown separately.}
    \label{fig:pr_response}
\end{figure}

\paragraph{ \textbf{(A3) Acceptance Ratios}}
\textit{Approach:} We identify whether or not there is a difference in how \ExPR s are accepted.
To do so, we classify the different state changes of a PR.
Similar to prior work \cite{Wang2021}, there are several states of a PR, that is \AcPR~-- where the PR has been closed and merged, and \AbPR~-- where the PR has been closed but has not been merged.
After identifying the state of each PR, we performed two analyses.
It is important to note that a PR can be either open or closed, merged or not merged.
Hence, since we are only concerned with acceptance (closed and merged) and abandonment (closed and not merged), we ignore other state (open and not merged).
Similarly to RQ1, we calculate the distribution of acceptance at the package level of analysis.
Hence, we use proportions as follows:
\begin{itemize}
    \item \texttt{state-\ExPR (x)}:~ $\frac{\# state-\ExPR}{\ExPR}$ that is submitted to a \texttt{pkg}. A higher proportion indicates that \ExPR s have a higher acceptance or abandonment rate for that package.
    \item \texttt{state-\InPR (x)}:~ $\frac{\# state-\InPR}{\InPR}$ that is submitted to a \texttt{pkg}. A higher proportion indicates that \InPR s have a higher acceptance or abandonment rate for that package.
    \item \texttt{state-\BotPR (x)}:~ $\frac{\# state-\BotPR}{\BotPR}$ that is submitted to a \texttt{pkg}. A higher proportion indicates that \BotPR s have a higher acceptance or abandonment rate for that package. 
\end{itemize} 
where \textit{x} refers to either an accept or abandon state.
Using these metrics, the first analysis is the distribution of the \ExPR s that were eventually accepted.
We test the following null hypothesis:
\begin{itemize}
    \item \HNullTwoOne
\end{itemize}
Similarly to P1 and P2, we use the Mann-Whitney test to test statistical significance.
We also measure the effect size using Cliff’s $\delta$, a non-parametric effect size measure \citep{cliff1993dominance}.
The effect size is analyzed as follows: 
(1) |$\delta$| $<$ 0.147 as Negligible, 
(2) 0.147 $\leq$ |$\delta$| $<$ 0.33 as Small, 
(3) 0.33 $\leq$ |$\delta$| $<$ 0.474 as Medium, and 
(4) 0.474 $\leq$ |$\delta$| as Large. We use the cliffsDelta\footnote{\url{https://github.com/neilernst/cliffsDelta}} package to analyze Cliff’s $\delta$.
For the second analysis, we compare the distributions between the three different PR types (\ExPR, \InPR, \BotPR).
For statistical validation, we test the following hypothesis:
\begin{itemize}
    \item \HNullTwoTwo~
\end{itemize}
First, we apply the Kruskal-Wallis H-test \citep{kruskal1952use},  which is a non-parametric statistical test to use when comparing more than two groups.
If there is significance, we then will conduct the Dunn test \citep{dunntest2015} to determine exactly which type of PRs are different to \ExPR s.
Additionally, we measure effect size using Cliff’s $\delta$.

\begin{table}[]
\centering
\caption{Summary statistics of each PR states of \ExPR, \InPR, \BotPR~i.e., Mean, Median, and SD.}
\label{tab:res_statistic}
\begin{tabular}{lcrrr}
\toprule
 \multicolumn{1}{c}{\multirow{2}{*}{\textbf{PR}}} & \multicolumn{3}{c}{\textbf{\% PR per pkg}} \\ \cline{2-4}
 &  \textbf{Mean}      & \textbf{Median}      & \textbf{SD}      \\ \midrule
 state-\ExPR (accept) & 55.65\% & 50.00\% & 33.88\% \\
 state-\InPR (accept) & 51.04\% & 50.00\% & 33.54\% \\
 state-\BotPR (accept) & 30.94\% & 22.73\% & 26.78\% \\ \midrule
 state-\ExPR (abandon) & 29.36\% & 20.00\% & 27.19\% \\
 state-\InPR (abandon) & 8.29\% & 5.31\% & 9.53\% \\
 state-\BotPR (abandon)& 31.18\% & 21.05\% & 28.10\% \\ 
\bottomrule
\end{tabular}
\end{table}

\begin{table}[!]
\centering
\caption{Statistical test of \ExPR~acceptance ratios}
\label{tab:res_statistical_ex}
\scalebox{0.9}{
\begin{tabular}{ccc}
\toprule
\multicolumn{2}{c}{\textbf{\ExPR}} & \textbf{Cliff’s $\delta$} \\ \midrule
\multicolumn{2}{c}{ state-\ExPR (accept) > state-\ExPR (abandon)*} & medium \\
\bottomrule
\multicolumn{3}{l}{Mann-Whitney U with *:p-value $<$ 0.001}
\end{tabular}}
\end{table}


\begin{table}[]
    \centering
\caption{Contingency table showing Dunn's test and Cliff's $\delta$ between \InPR, \ExPR, and \BotPR }
\label{tab:res_statistical_all}
\scalebox{0.9}{
\begin{tabular}{llll}
\toprule
 & \multicolumn{3}{c}{\textbf{Cliff’s $\delta$ (accept state | abandon state)}}                                    \\ \cline{2-4} 
 & \multicolumn{1}{c}{\ExPR} & \multicolumn{1}{c}{\InPR}          & \multicolumn{1}{c}{\BotPR} \\ \midrule
\multicolumn{1}{r}{\ExPR} & -                               & \multicolumn{1}{c}{negligible* | large*} & medium* | negligible       \\
\multicolumn{1}{r}{\InPR} &                                 & -                                        & medium* | large*           \\
\multicolumn{1}{r}{\BotPR}      &                                 &                                          & -                          \\ 
\bottomrule
\multicolumn{4}{l}{Dunn test *:p-value $<$ 0.001}                                                                                        
\end{tabular}
}
\end{table}

\textit{Results:}
Figure \ref{fig:pr_response} shows the acceptance ratios of all PR types per NPM package.
Note that the pink violin plot highlights the external PRs, split into accepted and abandoned rates.
Visually, we can see that both shapes are different. 
In terms of the accepted \ExPR s, we can see that the violin plot has a heavy top shape, indicating that there are many packages in the 90 to 100 percent ratio. 
On the other hand, looking at the abandoned ratio, we see that the violin plot is much thinner.
Complementary to the visual results, Table \ref{tab:res_statistic} shows that statistically, \ExPR s have a 55.65\% likelihood on average to be accepted. 
On the other hand, the chances for an \ExPR~to get abandoned is also lower, with a 29.36\% chance on average.
For statistical validation, Table \ref{tab:res_statistical_ex} confirms that there is indeed a statistical difference and confirms that \HNullTwoOne~ is accepted with a medium effect size.

Returning to Figure \ref{fig:pr_response}, we can also compare \ExPR s against the other PR types (i.e., \InPR~and \BotPR).
Interestingly, we can identify two findings.
The first is that the shape of the violin plots of \ExPR s and \InPR s are almost identical, with \BotPR s being different in terms of acceptance.
Complementing this result, Table \ref{tab:res_statistic} supports this, with both internal and external PRs having the same 50\% chance of being accepted. 
Different from this, the evidence suggests that \BotPR s are least likely to be accepted (i.e., 30.94\% on average). 
However, when it comes to abandonment, \ExPR s and \InPR s differ.
As shown in both the violin plot and the statistics in Table \ref{tab:res_statistic}, we see that developers are less likely to reject an \InPR~ (8.29\%) as opposed to an \ExPR~ (29.36). 
Similarly to \ExPR s, \BotPR s have the highest chance of being abandoned (31.18\%).
For statistical validation, Table \ref{tab:res_statistical_all} confirms that there is indeed a statistical difference and confirms that \HNullTwoTwo~ is accepted.
\begin{tcolorbox}
\textbf{Preliminaries}: 
\RqOneResult
\end{tcolorbox}
            
\section{Empirical Study}
\label{sec:empirical_study}
Motivated by the results of the preliminary study, we are now able to understand the need for \ExPR.

\subsection{Approach}
\textbf{To answer RQ1,} we conduct two analyses.
For the first analysis, we assume that developers' attention will be caught by PRs that resolve existing issues raised in a library.
Hence, we identify PRs that were created in response to an already existing issue that was raised in the library.
To do so, we link PRs to issues, following the process explained in the GitHub documentation\footnote{\url{https://docs.github.com/en/issues/tracking-your-work-with-issues/linking-a-pull-request-to-an-issue}}.
Concretely, we search for keywords (i.e., close, closes, closed, fix, fixes, fixed, resolve, resolves, resolved) in the description of the PR (cf. Section 2 for example). 
For our results, we will present statistics on the rate of PRs that can be linked to issues using proportions as follows:
\begin{itemize}
    \item \texttt{Linked External PR}:~ $\frac{\# Linked- \ExPR}{\#\ExPR}$ submitted to a \texttt{pkg}. In this case, a higher proportion means that there are more linked \ExPR s for that package. 
    \item \texttt{Linked Internal PR}:~ $\frac{\# Linked- \InPR}{\#\InPR}$ submitted to a \texttt{pkg}.  In this case, a higher proportion means that there are more linked \InPR s for that package. 
\end{itemize} 
Note that for this analysis, we decided not to report linked issues for Bots, due to a large amount of noise of keywords generated by the auto-generated messages that reference fixes outside of the project.



\begin{table}[]
\centering
\caption{Manually curated keywords of PR labels that `require attention'.}
\label{tab:critical_keyword}
\begin{tabular}{cl}
\toprule
\textbf{}            & \multicolumn{1}{c}{\textbf{identified keywords in labels}}                         \\ \midrule
Require attention    & attention, blocked, blocker, break, breaking,     \\
             & broken, change, confusing, critical, denied,  \\
             & density, difficult, difficulty, effort, exclamation,    \\
             & failed, failing, followup, hard, high, hold,            \\
             & important, incompatible, inconvenient, leak,     \\
             & memory, must, needed, notable, performance,  \\
             & prio, priority, required, risk, risky,  \\
             & securities, security, severity, urgent. \\
False positives & easy, low, medium, minor, no, non, review     \\
\bottomrule
\end{tabular}
\end{table}

For the second analysis, we assume that a PR might be labeled with words that might grab the attention of any core member, for example, a PR to fix a potential breaking change.
For this analysis, we first needed to filter out PRs that did not contain any labels. 
From the classified dataset, we ended up with 245,693 (78.18\%) \BotPR s, 26,797 (9.22\%) \ExPR s and 50,144 (14.73\%) \InPR s containing a label. Once we collected the PRs with labels, we performed three steps as part of pre-prossessing the labels.
In the first step, we remove non-textual symbols such as emojis, punctuation, and other non-English characters from the labels.
For the second step, we tokenize the label so that similar words can be grouped together.
Finally, in our last step, we apply lemmatization to derive the  common base form.
For the implementation, we used the Python packages \texttt{re}\footnote{\url{https://pypi.org/project/regex/}}  and \texttt{nltk}\footnote{\url{https://www.nltk.org/}}.

Table \ref{tab:critical_keyword} shows the curated list of keywords used to identify labels that would raise the attention of the core team.
Following related work \citep{mikaMSR17}, we identified words that would raise the attention (arousal) with words that show the priority levels of a label.
Using the base common words for each label, we then systematically and manually checked all unique common base labels to infer their importance.
The analysis involved two authors who manually scanned the words to classify keywords together, with one pointing out a label then the other mutually agreeing, similar to \cite{Subramanian2022}.
Furthermore, to remove false positives (such as the label `low severity'), we also collected a list of non-critical keywords.
We will report the percentage of these labels for all PR types (i.e., \ExPR, \InPR, and \BotPR).
Furthermore, we will report the number of PRs and also the top five labels within each type of PR.

\textbf{To answer RQ2,} we perform a qualitative analysis to characterize the differences in PR types between \InPR s, \ExPR s, and \BotPR s. 
In particular, we manually classify the sample PRs based on the taxonomy of \citet{Subramanian2022} as defined below:

\begin{itemize}
    \item \textit{Documentation}: Changes and additions made to documentation files such as READMEs and/or comments explaining code. Note: Only PRs, where the majority change is documentation, are classified as documentation change. 
    \item \textit{Feature}: Adding new functionality/features to the project. Following \cite{Subramanian2022}, we consider dependency updates as a feature change.
    \item \textit{Bug}: Fixing unexpected behaviour in code.
    \item \textit{Refactoring}: Restructuring code to make it more understandable/readable and/or conform to coding standards.
    \item \textit{GIT related issues}: Solving merge conflicts, adding elements to .gitignore files and other changes related to GIT.
    \item \textit{Test cases}: Adding test cases and/or adding code to facilitate testing.
    \item \textit{Other}: Anything that does not fall into the above categories.
\end{itemize}


We conducted manual coding to classify our sample PRs, a technique that is popular in various qualitative studies of software engineering \citep{wattanakriengkrai2020github}. 
First, three authors of this paper independently coded the PR types for 45 samples (i.e., containing 15 from \ExPR, \InPR, and \BotPR~ in a first iteration. 
They consider the PR title, description, conversation, and code changes.
We then calculated the inter-rater agreement between the categorization results of the first three authors using Cohen's kappa \citep{mchugh:2012kappa}.
The kappa agreement varies from 0 to 1 where 0 is no agreement and 1 is perfect agreement.
This process is iterated until there is a kappa agreement of more than 0.8 or ``almost perfect'' \citep{Viera:2005}.

For this analysis, we obtain a kappa agreements of 0.91 for overall (\ExPR, \InPR, and \BotPR), 0.95 for \ExPR, 0.90 for \InPR, and 0.95 for \BotPR~ in the second round of iterations.
After this high agreement score, the remaining data in the samples of 384 \InPR s, 384 \ExPR s, and 384 \BotPR s is divided into three parts, and each part is coded separately by three authors of the paper.
We apply Pearson’s chi-squared test ($\chi^2$) \citep{Pearson1900} to test whether content of PR to \ExPR~and \InPR~are independent or not. 
To show the power of differences between each \ExPR~and \InPR, we investigate the effect size using Cram\'er's V ($\phi'$), which is a measure of association between two nominal categories \citep{cramer2016mathematical}.
According to \citet{Cohen:1988}, our grouping has one degree of freedom (df*), hence, effect size is analyzed as follows: (1) $\phi'$~\textless~0.07 as Negligible, (2) 0.07~$\le$~$\phi'$~\textless~0.20 as Small, or (3) 0.30~$\geq$~$\phi'$~as Large.

\subsection{Findings}
\label{sec:findings}
In this section, we present our findings to answer our two research questions.

\begin{table}[t]
\centering
\caption{Summary statistic of the linked \ExPR~and linked \InPR~i.e., Mean, Median, and SD (\RqThree)}
\label{tab:linked_pr}
\begin{tabular}{ccccc}
\toprule
\multirow{2}{*}{\textbf{PR}} & \multirow{2}{*}{\textbf{\#linked PR}} & \multicolumn{3}{c}{\textbf{\% linked-PR per pkg}} \\ \cline{3-5} 
 & & \textbf{Mean}      & \textbf{Median}      & \textbf{SD}      \\ \midrule
Linked-\ExPR& 22,662 & 26.75\% & 16.67\% & 25.68\% \\
Linked-\InPR & 35,980 & 42.57\% & 33.33\% & 32.93\%     \\    
\bottomrule
\end{tabular}
\end{table}


\begin{table*}[]
\centering
\caption{Ratio of PRs that were labeled as `require attention'.}
\label{tab:ratio_critical_label}
\begin{tabular}{lrrr}
\toprule
\textbf{} & \multicolumn{1}{c}{\textbf{\# PR}} & \multicolumn{1}{c}{\textbf{\# PR}} & \multicolumn{1}{c}{\textbf{\# attention}} \\
\textbf{} & \multicolumn{1}{c}{\textbf{with label}} & \multicolumn{1}{c}{\textbf{attention label}} & \multicolumn{1}{c}{\textbf{label}} \\ \midrule
\ExPR & 26,797 & 984 (3.63\%) & 166 (61.25\%)   \\  \hline
\InPR & 50,144 & 2,177 (4.34\%) & 205 (75.65\%) \\ \hline
\BotPR & 245,693 & 3,392 (1.38\%) & 26 (9.59\%) \\ \bottomrule
\multicolumn{1}{r}{Total}  & 322,634 & 6,553 (2.03\%) & 271 (100\%) \\
\end{tabular}
\end{table*}


\textit{{\uline{PRs linked to existing issues}}}.
In Table \ref{tab:linked_pr}, we report the percentage of linked \ExPR s.
We find that internal PRs are more likely to be linked with an average of 42.57\% per package.
This result may not be surprising, as we assume that any core team would be focused on resolving any existing issues in their projects.
Interestingly, on average 26.75\% of \ExPR s per package are also linked. 
This evidence shows that a quarter of \ExPR s were in response to an existing issue.

\textit{{\uline{Labels that require attention}}}.
Table \ref{tab:ratio_critical_label} shows the frequency count of PRs that might raise developer attention.
As expected, the number of these PRs should not be large, as they indicate critical problems for a library.
Yet, when looking at the percentages, we deduce that although small, \ExPR s do have relatively comparative ratios (3.63\%) compared to \InPR s (4.34\%) and \BotPR s (1.38\%).
It is important to note that since labeled PRs account for a small portion of PRs (i.e, 245,693 (78.18\%) \BotPR s, 26,797 (9.22\%) \ExPR s and 50,144 (14.73\%) \InPR s), at this stage we cannot make any statistical generalizations.
 Instead, the evidence shows that \ExPR s contain PRs that require the attention of developers and might provide fixes to critical issues. 

\begin{table*}[]
\centering
\caption{Top five attention unique labels of each PR type.}
\label{tab:top_five_critical_label}
\begin{tabular}{ll}
\toprule
\textbf{} & \textbf{Top five attention unique labels (\#)} \\ \midrule
\ExPR & breaking (78), P3: important (50), P1: urgent (46), on hold (45), \\
      &  Breaking Change (40) \\ \hline
\InPR  & breaking (311), breaking change (141), P2: required (104),  \\
      & change/patch (96), performance (80)  \\\hline
\BotPR  & security (3248), breaking (28), CH: Security (23),  \\
      & Type: Security (21), Security (11)  \\ \bottomrule
\end{tabular}
\end{table*}


\begin{table*}[!th]
\caption{Examples of PR content with title and reason for classification.}
\label{tab:example_type_contribution}
\begin{tabular}{lll}
\toprule
\textbf{PR Content}  & \textbf{PR title} & \textbf{Reason for classification}\\ \midrule 
Documentation & Update WritingTests.md & The PR modifies the file named \\
    & \citep{onlineDoc} & `WritingTests.md' \\
Feature & feat: Add twitch icon \citep{onlineFeat} & PR's title includes `feat' keyword.\\
Bug & BIG-21501 - EU Cookie warning & PR's title and description contain\\
    & (Bugfix) \citep{onlineBug} & `bugfix' keyword.\\
Refactoring & Major refactoring \citep{onlineRefactor} & PR's title includes `refactoring' \\
     &   & keyword.\\
GIT related & Merging cards theme into master & The PR is for merging another\\
issues & \citep{onlineGitRelated} & branch to master.\\
Test cases & Remove TLS account creation & PR's description explains about \\
     & tests \citep{onlineTestCases} & test integration.\\
Other & Mark the package as having & The PR modifies a config file\\
     & no side effects \citep{onlineOther} & (package.json)\\
\bottomrule
\end{tabular}
\end{table*}

In terms of the content of the PRs that require the attention of developers, as shown in Table \ref{tab:top_five_critical_label}, the top five unique labels for \ExPR~include breaking (78), important (50), urgent (46), on-hold (45) and breaking change (40). 
Differently \InPR~ includes labels such as required (104), change/patch (96), and performance (80), which \BotPR~include labels related to security.
Interestingly, \BotPR s seem to specialize in fixing security issues, while only having 9.59\% of the total unique label keywords.
This result is consistent with the intuition that many of the bots detected were dependency management tooling, which may include security vulnerability fixing.
Based on these results, we now return to answer the first research question \textbf{\RqThreeText}
\begin{tcolorbox}
\textbf{Summary for \RqThree}: 
\RqThreeResult
\end{tcolorbox}


\textit{{\uline{Most \ExPR~relate to New Features}}}.
Table \ref{tab:example_type_contribution} shows examples of each PR change type, while Figure \ref{fig:sample_contribution_type} depicts the frequency count for each type in our sampled dataset.
A key finding of our result is that most PRs were related to creating a new feature, which is consistent with the results of \citet{Subramanian2022}.
In terms of types, \BotPR s (380 out of 384 PRs) were the most frequent for bots and also for the category.
The most likely reason is that the NPM ecosystem has been popular for using bots to assist with dependency management.
Libraries in the ecosystem have been the target of popular bots such as dependabot, greenkeeper, and other dependency management bots. 
This is confirmed by related work \citep{Rombaut2022}.

\pgfplotstableread[col sep=comma,]{data.csv}\datatable

\begin{figure*}
\centering
\begin{footnotesize}
\begin{tikzpicture}[scale=.9]
\begin{axis}[
    align=center,
    xbar= 0.0cm,
    bar width=10pt,
    ytick=data,
    yticklabels from table={\datatable}{A},
    height = 10cm,
    legend pos=south east,
    axis line style={opacity=0},
    major tick style={draw=none},
    nodes near coords,
    nodes near coords align={horizontal},
    point meta=rawx,
    enlarge x limits = {value = 0.25, upper},
    ]
    \addplot [fill=green!40,draw=none] table [y expr=\coordindex, x=Internal]{\datatable};
    \addplot [fill=pink,draw=none] table [y expr=\coordindex, x=External]{\datatable};
    \addplot [fill=cyan!40,draw=none] table [y expr=\coordindex, x=Bot]{\datatable};
    
    \legend{Internal, External, Bot}
\end{axis}
\end{tikzpicture}
\end{footnotesize}
\caption{Frequency count of seven PR types comparing \InPR, \ExPR, and \BotPR}
\label{fig:sample_contribution_type}
\end{figure*}
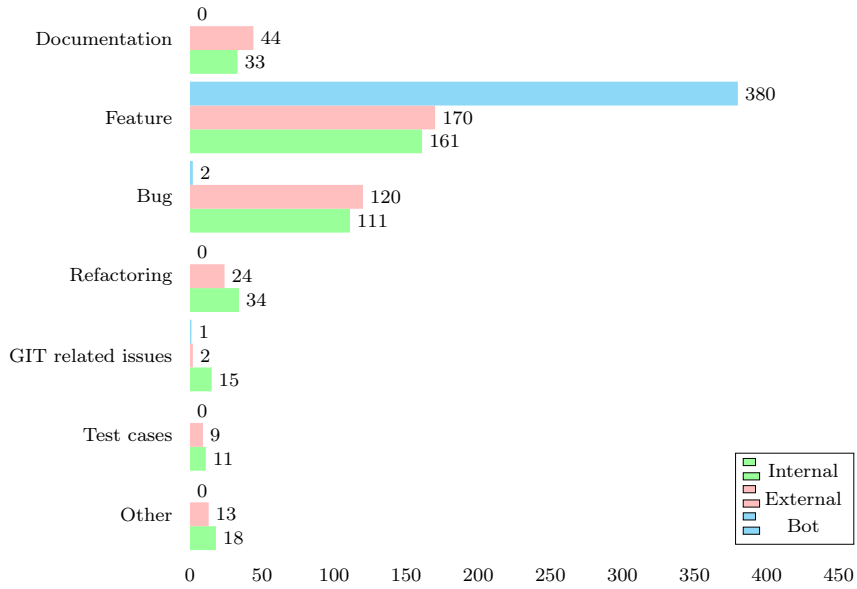

\textit{{\uline{\ExPR~also relate to features, bugs, and documentation, but not refactorings or GIT related issues}}}.
When comparing between the different types of PRs, we find that \ExPR s target features (170 PRs), bugs (120 PRs), and documentation (44 PRs).
\InPR s similarly add features (161 PRs), fix bugs (111 PRs), but in addition, have more refactoring-related PRs (34 PRs), and GIT-related issues (15 PRs). 
We find that statistically, \ExPR~and \InPR~have different PR contents. (i.e., $\chi^2 = 14.837$, \textit{p-value} $< 0.021$). Cram\'er's V effect size ($\phi'$), indicated a small level of association.
We now return to answer the final research question \textbf{\RqFourText}
\begin{tcolorbox}
\textbf{Summary for \RqFour}: 
\RqFourResult
\end{tcolorbox}

\section{Lessons Learnt}
\label{sec:recommendations}
Returning to the goal of the study, we find that \ExPR s are almost as important as Internal PRs submitted by members of the core team.
Hence, we make the following recommendations and highlight challenges and future work that this work sparks.

\begin{itemize}
    \item \textit{Libraries Need External PRs.}
    Based on the preliminary study, RQ1 and RQ2, we find evidence that the ability of a library to attract and sustain External PRs is crucial.
    We find that most packages receive External PRs, and an ecosystem contributor is likely to submit External PRs.
    Thus we show that it is important for library packages to have a constant flow of external contributions.
    This might be even more important in cases where a library has a single maintainer.
    Hence, we suggest that libraries consider strategies to attract the community to their library.
    One interesting avenue for future work would be exploring the motivations for why an external submitter would make a contribution and what factors might attract them to make a contribution.

    \item \textit{External PRs attract the attention of maintainers.}
    RQ1 shows that \ExPR s~can contain changes that attract the attention of the core team. 
    This is good news, as this assistance might provide new ideas or lend a hand to often overworked maintainers. 
    With just over a quarter of \ExPR s~being linked to an already existing issue, this means that these PRs are directly related to actionable and existing problems that can be immediately reviewed and merged into the codebase.
    Potential research directions would be understanding those External PRs that are not linked.
    Also, in this study, our evidence only comes from mining the artifacts; the next logical step would be to conduct interview or survey studies to confirm that maintainers see these External PRs as assisting them in maintenance, rather than adding another layer of work for them.
    Another future research direction would be to explore whether the continued supply of different types of External PRs (RQ2) is a positive factor for the sustainability of a library (to avoid becoming obsolete).

    \item \textit{External PRs meet documentation needs.}
    Results from RQ2 show that External PRs meet needs that are different from Internal PRs, i.e., they are more likely to help with documentation and feature requests.
    This is especially good news for very active projects such as software libraries that require constant updates to fix bugs and add new features. 
    It is good for maintainers to know that there is a community that helps them manage new features, deprecation, and other critical aspects that need to be communicated back to their users.
    A potential research direction is to investigate how these documentation needs can be met by automated approaches, to further reduce the workload of maintainers.
    These results might also indicate that creating an External PR might not require much technical skill or coding experience, as the focus can be on updating the documentation or correcting inconsistencies in its usage.
    
    On the other hand, we find that External PRs are less likely to submit refactoring changes for a project.
    This is intuitive, as performing such operations requires more in-depth knowledge of the code.

\end{itemize}

\section{Threats to Validity}
\label{sec:threats_to_validity}
\textit{Internal validity} - We discuss three internal threats.
The first threat is the correctness of the techniques used in the mining and extracting of our datasets.
As we use the listed NPM packages and pull request information based on \citet{chinthanet2021lags}, we are confident that our result can be replicated. 
Furthermore, we contribute a full dataset that can be used by future researchers. 
The second threat is related to tool selection (e.g., statistical testing).
This is because different data sources and tests may lead to different results.
To mitigate this threat, we use standard tests such as Spearman's rank correlation coefficient, Mann-Whitney U test, and Kruskal-Wallis H-test for statistical validation from popular and well-known Python packages (i.e., SciPy Statistical functions\footnote{\url{https://docs.scipy.org/doc/scipy/reference/stats.html}} (scipy.stats))
The third threat is through our qualitative analysis or manual classification in \RqFour.
We mitigate these threats by reporting the agreement among three raters using Cohen's kappa.
In this case, all three raters discussed refining the types of PR content until they achieved a high kappa agreement (more than 0.8).

\textit{Construct validity} - Threats to the construct validity of this work are related to the categorizations and assumptions in the study.
A threat is the identification of \ExPR s, which is prone to false positives. 
Different from prior work by \citet{Valiev2018sustained}, we define a core contributor as a contributor who has the ability to merge a PR.
The bot classification is based on prior approaches by \citet{dey2020detecting} and \citet{golzadeh2020bot}.
Furthermore, for \RqThree, our identification of keywords is prone to generalization issues.
We acknowledge these threats, but are confident as we carefully use empirical standards for systematic classification, with all data available for replication. 
Finally, we do not take into account the internal activities of maintainers. 
We believe that this is outside the scope of this work, however, will be considered for future work. 

\textit{External validity} - The main threat to external validity exists in the generalizability of our results to other package ecosystems.
In this study, we focused solely on the NPM JavaScript ecosystem which has around 2 million packages, with over 180 million downloads in July 2022.
However, we have no reason to believe that our analysis is not applicable to other ecosystems that have similar package management systems, e.g., PyPI for Python.
Although we analyzed a large number of repositories on GitHub, we cannot generalize our findings to industry or open-source repositories in general. 
Some open-source repositories are hosted outside of GitHub, e.g., on GitLab or private servers.
To mitigate threats to \textit{reliability}, we are careful to state that these recommendations and implications may only be specific to the NPM community, and extension is seen as immediate future work.

\section{Related work}
\label{sec:related_work}
Our work is situated between two research fields.
We now introduce and highlight related literature.

\paragraph{\textbf{Studies on Third-party Libraries}}
Third-party libraries have been studied in various aspects, including third-party library reuse \cite{Heinemann:reuse2011, abdalkareem2017developers, Xu:emse2020}, third-party API library \cite{ALRUBAYE2020,thung2016api}, library migration \cite{HeHao2021,Cogo2019}, \cite{huang2020interactive}, and security vulnerabilities in third-party libraries \cite{chinthanet2021lags,decan2018impact}.
Only a few studies have focused on understanding developer contributions to third-party libraries, which we believe plays an important role in library sustainability.
In terms of sustainability of packages, \citet{dey2019_promise_mockus} observed that users contribute and demand effort primarily from packages that they depend on directly with only a tiny fraction of contributions and demand going to transitive dependencies, with a case study of NPM packages.
Recently, \citet{Dey:ESEM2020} revealed the significant effects of technical and social factors on the developer contribution quality for NPM packages. Although these papers are complementary, to the best
of our knowledge, the characteristics of external contributions for third-party libraries have not been studied comprehensively.
Differently, \textbf{in our work}, we focus on external contributions to third-party libraries, looking at their prevalence and their characteristics.

\paragraph{\textbf{Studies On OSS Sustainability}}
To sustain OSS projects, previous studies have extensively investigated the motivations and barriers to developers’ joining and retention.

\textit{Motivations to make OSS contributions} - The common motivations to make contributions to OSS projects are the joy of programming, the identification with a community, career advancement, and learning \cite{Hars:2001}.
Additionally, \citet{Roberts:2006} explored the interrelationships between motivations of OSS developers, revealing that motivations are not always complementary.
When comparing motivations between individual developers and companies, \citet{Bonaccorsi:2006} observed that companies are more motivated by economic and technological reasons.
\citet{Lee:icse2017} found that the most common motivation of one-time contributors is to fix bugs that affect their work, while the highly mentioned motivation of casual contributors is ``scratch their own itch'' \cite{Pinto:saner2016}.
\textbf{In our work}, although we do not study the motivations for contributing to a library, our study is related to characterizing OSS contributions.

\textit{Barriers to participation in OSS projects} - A study by \citet{Fagerholm:esem2014} observed that mentoring increases the chance of developers’ active participation.
In the pull-based model, a survey by \citet{Gousios:icse2016} revealed that the most commonly reported challenge is the lack of responsiveness of project integrators.
\citet{Steinmacher:chase2013} investigated the first interactions of newcomers in an OSS project. Their results showed that these interactions affected the onboarding of newcomers.
Furthermore, they analyzed pull requests of quasi-contributors and found that non-acceptance demotivated or prevented them from placing another contribution \cite{Steinmacher:icse2018}.
\citet{Assavakamhaenghan:msr2021} tried to understand the correlation between the first response given to the first contribution and future contributions.
\citet{Li:tse2021} observed the significant impacts of pull request abandonment on project sustainability.
Complementary, \textbf{in our work}, we also analyze pull requests but with a focus on third-party libraries rather than more generic OSS projects.

\textit{Developers’ retention and engagement in OSS projects}
A study by \citet{Zhou:icse2012} found that developers’ willingness and participation environment significantly impact the chance of becoming long-term contributors.
In addition, \citet{Schilling:hawaii2012} showed that developers’ retention in OSS projects is affected by the level of development experience and conversational knowledge.
\citet{Valiev2018sustained} conducted a mixed-methods
study to investigate ecosystem-level factors that affect the sustainability of open-source Python projects. Their results show that projects with more contributors are less likely to become dormant.
\citet{Iaffaldano:IEEE2019} conducted interviews with OSS developers to explore what drives them
to become temporarily or permanently inactive in a project. The reported reasons were personal (e.g., life events) or project-related (e.g., role change and changes in the project).
Recent studies have also focused on understanding what contributions attract newcomers \citep{Subramanian2022}, \citep{Rehman:2020}.
Meanwhile, the usefulness of automated pull requests, i.e., bots, has attracted interest from researchers in these research areas.
Through an empirical study of OSS projects with bots,
developers revealed that these bots are useful for maintaining projects \citep{Wessel:2018}.
\citet{Mirhosseini:IEEE2017} observed that bots can encourage developers to contribute to OSS projects, especially to update dependencies.
\citet{alfadel2021botPR} studied Dependabot, a bot used to automatically update vulnerable dependencies. They found that 65\% of the created security-related pull requests are accepted.
\textbf{In our work,} we also observed PRs submitted by bots, finding that bots such as Dependabot are prevalent in \BotPR s.

\section{Conclusion}
\label{sec:conclusion}
This work investigates the extent to which third-party libraries receive support from external contributors (i.e., \ExPR s) that are not part of the core team.
By mining PRs and analyzing them through mixed methods of statistical analysis and qualitative analysis, we show that \ExPR s are prevalent and just as likely to be accepted as \InPR s.
Our results reinforce the call of core teams that need support.
Future work includes research into encouraging and sustaining \ExPR s, understanding how \ExPR s could be automated, or could be the focus for newcomers to the ecosystem.
We envision that this work can inspire awareness and initiatives to support third-party libraries.

\section*{Acknowledgement}
This work is supported by Japanese Society for the Promotion of Science (JSPS) KAKENHI Grant Numbers 20K19774 and 20H05706.

\section*{Statements and Declarations}
Raula Gaikovina Kula and Christoph Treude are members of the EMSE Editorial Board.

\bibliographystyle{spbasic}      
\typeout{}
\bibliography{bibliography}   

\end{document}